\documentclass[11pt]{article}
\usepackage[dvips]{graphicx}
\def\Vec#1{\mbox{\boldmath $ #1 $}}

\begin{document}
\begin{center}
\begin{LARGE}
Model-independent derivation of macroscopic Maxwell equations from microscopic basis: 
Beyond the "$\epsilon$ and $\mu$ " description  \\[1cm]
\end{LARGE}
Kikuo Cho \\ 
Toyota Physical and Chemical Research Institute, 
Nagakute, Aichi 480-1192, Japan \\[1cm]
\end{center}

\begin{abstract}

Pointing out the incompleteness of conventional macroscopic Maxwell equations 
(M-eqs.), we propose a new form derived from the long wavelength approximation (LWA) 
of microscopic nonlocal response.  From the general Hamilonian of matter and 
matter-EM field interaction (containing spin dependent terms due to relativistic 
correction), we first set up the simultaneous equations for microscopic "vector 
potential $\Vec{A}$ and induced current density $\Vec{I}$", and then extract 
the macroscopic components by applying LWA.  This leads to new macroscopic M-eqs. 
with a single macroscopic susceptibility $\chi_{\rm em}(\Vec{k}, \omega)$ between 
$\Vec{I}$ and $\Vec{A}$, which describes both electric and magnetic polarizations 
and their mutual interference in its fully quantum mechanical expression.  In the 
absence of chirality and under the condition to use magnetic susceptibility defined 
with respect to $\Vec{B}$, this scheme is shown to be equivalent to the conventional 
"$\epsilon$ and $\mu$ scheme".  In the case of chiral symmetry, the phenomenological 
constitutive equations by Drude, Born and Fedorov cannot be justified by this 
microscopic approach.  As a single susceptibility scheme of macroscopic M-eqs., 
this result is on an advanced level by its fully quantum mechanical description 
of the whole set of $O(k^0)$, $O(k^1)$ and $O(k^2)$ terms of susceptibility 
providing a consistent picture of such a scheme. 
\end{abstract}

\section{Introduction} 


Maxwell equations (M-eqs.) are one of the most fundamental theoretical frameworks 
of physics, dealing with the interaction between electromagnetic (EM) field and 
charged particles with magnetic moment, which describes the EM response in both 
static and dynamical regimes.  They consist of four laws, i.e., Gauss laws for 
electricity and magnetism, Ampere's law and Faraday's law.   There are two forms 
of them, i.e., microscopic and macroscopic M-eqs., and both are actively used for 
research and education.   The fundamental formulas and definitions about micro- 
and macroscopic M-eqs. are given with usual notations in Appendix A.  

Historically, macroscopic M-eqs. appeared first as a consistent set of the four 
equations mentioned-above for macroscopic matter systems.  It was a phenomenology, 
since nothing was known about electrons, quantum mechanics and relativity at that 
time.  As the particle picture of matter was established, one could correctly 
transform macroscopic M-eqs. to microscopic ones describing the EM field in 
microscopic systems \cite{Lorentz}.  The microscopic form, combined with quantum 
mechanics and relativity, lead to quantum electrodynamics (QED), which is the most 
accurate theory of physics in precise agreement with experiments 
\cite{Cohen-Tann, QED2}.  

On the other hand, the particle picture of matter in combination with quantum 
mechanics has been used to promote macroscopic M-eqs. from a phenomenological level 
to a semiquantitative one by treating susceptibilities,  not as parameters, but as 
the quantities to be calculated quantum mechanically.  It is based on the viewpoint 
that macroscopic behavior of EM field must somehow be derived from the averaged 
motions of microscopic EM fields and charged particles.  This type of rewriting has 
been called the {\it derivation} of macroscopic from microscopic M-eqs., and such 
semiquantitative M-eqs. in terms of the field variables "$\Vec{E}, \Vec{D}, \Vec{B}, 
\Vec{H}$" have been used with much success for the studies of 
macroscopic EM response in a vast area. 

However, in spite of the long history with great success, the macroscopic scheme 
in terms of "$\Vec{E}, \Vec{D}, \Vec{B},\Vec{H}$" seems to be incomplete in comparison 
with microscopic one with respect to the follwing points. 
\begin{enumerate}
\item [1)] Lack of explanation why the number of susceptibility tensors are 
           different betweeen microscopic and macroscopic M-eqs. 
\item [2)] Lack of uniqueness in dividing charge density into true and 
           polarization charge densities 
\item [3)] Lack of uniqueness in dividing transverse current density 
           into the contributions of magnetic and electric 
           polarizations
\item [4)] Different wave number dependence of magnetic permeability $\mu$ 
           for two types of magnetic dipole (M1) transitions, i.e., spin 
           and orbital M1 resonances
\end{enumerate}
The details of each point are explained below. 
\begin{enumerate}
\item[1)]
In the derivation of macroscopic M-eqs. from microscopic ones, which can be 
found in many textbooks of electromagnetism \cite{EMtext}, it is admitted from 
the beginning that the two vector fields, i.e., electric and magnetic 
polarizatoins, are the fundamental quantities of macroscopically averaged matter.  
This is, however, not obvious, because in the more basic microscopic M-eqs. 
the dynamical variable of matter for EM response is a single vector field, i.e., 
"current density".  (Charge density is not an independent variable, since it is 
related with current density via continuity equation.)  This means that the 
number of constitutive equations (or susceptibilities) is different between 
the microscopic and macroscopic M-eqs..  It is not obvious that macroscopic 
averaging requires such an increase in the number of constitutive equations  
\cite{ChoICPS}.

 For systems with chiral symmetry, where elecrtic field can induce elecrtic, as 
well as magnetic polarization, (and magnetic field can induce magnetic, as well 
as electric polarizations,) a phenomenological approach by Drude-Born-Fedorov 
(DBF) constitutive equations is known to be useful \cite{chiral}, \cite{Band}.  
(See Appendix A.)  This scheme requires an additional (chiral) susceptibility.  
Then, the number of suseptibility in macroscopic M-eqs. is increased to three, 
in further contrast with microscopic M-eqs. 

\item[2)]
It is obvious that the separation of a given charge density $\rho(\Vec{r},t)$ 
into true and polarization charge densities, $\rho_{\rm t}(\Vec{r},t)$ and 
$\rho_{\rm p}(\Vec{r},t)$, respectively, cannot be done uniquely.  Since the 
only constraint is the charge neutrality $\int{\rm d}\Vec{r}\ \rho_{\rm p} = 0$, 
there are infinitely many ways to rewrite $\rho$ into $\rho_{\rm t} + 
\rho_{\rm p}$.  Because of the continuity equation, this problem can be 
rephrased as the non-uniqueness of separating the longitudinal (L) component of 
current density $\Vec{J}$ into the conduction current density $\Vec{J}_{\rm c}$ 
and the L component of the current density $\partial \Vec{P}_{\rm L}/\partial t$ 
induced by electric polarization $\Vec{P}$. 
\item[3)]
A similar problem exists in separating the transverse (T) component of $\Vec{J}$ 
into magnetization ($\Vec{M}$) induced current density $c\ \nabla \times \Vec{M}$ 
and the T component of $\Vec{P}$-induced current density 
$\partial \Vec{P}_{\rm T}/\partial t$. 
In order to do this division uniquely, we need to define $\Vec{P}$ and $\Vec{M}$ 
precisely, which, however, is not always possible. Beause in chiral symmetry 
the electric dipole (E1) and magnetic dipole (M1) transitions are mixed, which 
allows $\Vec{P}$ induced by magnetic, as well as electric field and $\Vec{M}$ 
induced by electric, as well as magnetic field.   Then, there is no unique 
way to separate $\Vec{P}$ and $\Vec{M}$. 

\item[4)]
The problem about $\mu(\omega)$ is the absence of its general expression including the 
contributions from both spin resonance and orbital magnetic dipole (M1) transitions.  
Since the deviation of $\mu(\omega)$ from unity is due to the M1 transitions of 
matter system, one would expect similar forms of resonant terms for both transitions. 
However, the conventional ways to write the two contributions 
are different with respect to their dependence on the wave number of the resonant EM waves. 
In the case of spin resonance \cite{Slichter}, a typical magnetic susceptibility has the form 
$\chi_{\rm m} = b/[\omega_{0} - \omega - i \gamma]$, where $\omega_{0}$ is the transition 
energy of a spin resonance, and $b$ the squared transition matrix element of spin 
magnetization operator.  The resultant expression $\mu = 1 + 4\pi \chi_{\rm m}$ does not 
contain the wave number of the resonant EM field except for $\omega$ in the denominator. 
On the other hand, M1 component of orbital transitions, which occur in any energy regions 
including nuclear transitions (e.g., M\"ossbauer transitions), produces an EM field with 
an amplitude proportional to the product of the resonant wave number $k_{\rm r}$ and the 
magnetic dipole moments of the transition, according to the standard multipole expansion 
of the EM field \cite{M1-tr}.  This means that the $\mu(\omega)$ in this case has 
a resonant term of intensity proportional to $k_{\rm r}^2$.  Thus, the two cases of M1 
transition apparently show different $k_{\rm r}$ dependences of $\mu(\omega)$. 

  Another difficulty about $\mu$ is that, in the standard dispersion equation 
$(ck)^2/\omega^2 = \epsilon \mu$, there appears the product, not the sum, of electric 
and magnetic contributions in spite of linear response treatment.   If we consider 
the possibility of the mixing of electric and magnetic excitations in chiral symmetry, 
there is no essential distinction between them as matter excitations.  Therefore, it 
would be more reasonable to have their contributions as a sum, rather than as a product, 
in the dispersion relation of linear response.   Though this problem looks different from 
the one in the previous paragraph, it turns out later that they have a common origin. 
\end{enumerate}

To the author's understanding, the origin of the incompleteness mentioned above 
lies in the use of $\Vec{P}$ and $\Vec{M}$ as the independent, fundamental variables 
in deriving macroscopic M-eqs.  This independence is questionable, if we think of 
chiral symmetry where electric and magntic transitions contributing to $\Vec{P}$ 
and $\Vec{M}$ are mixed with each other.  As a logic to derive macrosacopic M-eqs. 
in the conventional manner, it is necessary to show the general definition 
of $\Vec{P}$ and $\Vec{M}$ from the first principles, i.e., from the precisely defined 
matter Hamiltonian and matter-EM field interaction.   As will be discussed in Sec. 3.1, 
one would realize the incompleteness from the very effort to derive the matter-EM field 
interaction in an appropriate form to calculate electric and magnetic susceptibilities 
from the first principles. 
  
Therefore, in order to find right answers to the problems mentioned above, we need 
to use a new and more reliable logic in deriving macroscopic M-eqs.  As such, we start 
from a microscopic response theory, a most reliable basis for the EM response of matter 
in semiclassical regime.  In this scheme, all the details of matter states are 
described quantum mechanically, and all the wavelength components of EM field are taken 
into account in a semiclassical way.  Matter Hamiltonian and matter-EM field interaction  
are precisely defined from the Lagrangian for a general interacting system of charged 
particles and EM field, and explicit spin dependence is considered as relativistic 
corrections.  In the microscopic response theory, we solve, so to speak, Schr\"odinger 
eq. and micoscopic M-eqs. selfconsistently, or more specifically, the coupled integral 
equations for induced current density and vector potential, which contain all the 
wavelength components.   This way of solution becomes feasible due to the separability 
of the microscopic nonlocal susceptibility as an integral kernel, as will be shown in 
Sec.2.  We call this scheme "microscopic nonlocal response". 

Then, as a straighforward procedure of macroscopic averaging, we apply long wavelength 
approximation (LWA) to the coupled integral equations.  Thereby, we do not assume or 
request to recover the conventional form of macroscopic M-eqs. but just take the outcome 
as a logical consequence of this approach and compare the result with conventional form. 
Here also, the separability of microscopic susceptibility plays an essential role to 
give an explicit form of macroscopic susceptibility quantum mechanically.  This type 
of derivation has never been done before to the author's knowledge, and in fact it gives 
reasonable answers to all the problems mentioned above.  It should be noted that this 
approach describes the macroscopic averaging, not just as a concept which might allow 
various interpretations, but as an unambiguous mathematical procedure of LWA applied to 
the well-defined integral equations of microscopic nonlocal response.   This enables 
us, in principle, to check the validity condition for the approximation, since we 
clearly know what is discarded by LWA.  This point will be discussed in Sec.3.5. 

The most remarkable result is that the M-eqs. remain in the same form 
after applying LWA, and the only difference between the micro- and macroscopic M-eqs. 
lies in the form of the susceptibility.   Among several proposals of the single 
susceptibility scheme of macroscopic M-eqs. \cite{AgranoSpd, Agrano, Keldysh}, 
the present one is on 
an advanced level in the sense that all the components of susceptibility are given in 
a qunatum mechanical way, which, in principle, allows microscopic evaluation, and that 
it is a logical consequence from the more basic microscopic response scheme.  Detailed 
discussions will be given in Sec.3.2.   

The existing schemes of M-eqs. in the semiclassical regime are classified into the 
following hierarchy in the order of decreasing accuracy: (i) microscopic nonlocal 
response, (ii) the new macroscopic scheme, (iii) the conventional "$\epsilon$ and $\mu$" 
scheme.   Via LWA, (i) reduces to (ii), and (iii) is obtained from (ii) in the absence 
of chirality and by the use of magnetic susceptibility $\chi_{\rm B}$ defined with 
respect to $\Vec{B}$.  The justification of the step from (ii) to (iii) goes in a clear 
cut way as shown later, but a similar argument is not possible for the justification of 
the DBF equations for chiral symmetry, so that the DBF eqs. should stay outside the 
hierarchy as a phenomenology. 

The use of $\chi_{\rm B}$ is, not only a logical consequence, but also the solution of 
the problem 4).  Rewriting the dispersion equation $(ck)^2/\omega^2  = \epsilon \mu$ 
in terms of $\mu = 1/(1 - 4\pi \chi_{\rm B}$), obtained from $\Vec{M} = \chi_{\rm m}\Vec{H} 
= \chi_{\rm B}\Vec{B}$ and $\mu = 1 + 4\pi \chi_{\rm m}$, we immediately see the solutions 
for the both problems of 4), which will be discussed in Sec.2.4 in more details.   
The requirement to use $\chi_{\rm B}$ will be a considerable impact on the study of LHM.  
In Sec.3.4 an observable difference is demonstrated in the forms of dispersion curve 
and reflectivity spectrum in a LHM situation for the cases of assigning magnetic excitation 
energy to the pole of $\chi_{\rm m}$ and $\chi_{\rm B}$. 

Macroscopic M-eqs. are still an important tool in many research fields, such as 
photonic crystals \cite{PC}, left-handed materials (LHM) or metamaterials \cite{LHM}, 
near field optics \cite{NFO}, and so on.  For this reason and also for its importance 
in the elementary course of physics education, it is worth looking for a more complete 
form of macroscopic M-eqs.  Especially the present investigation, which raises 
a question \cite{ChoICPS} about the validity of the  Veselago's condition for LHM 
"$\epsilon < 0, \mu < 0$ \cite{Veselago}, has direct relevance to the today's very 
popular subject, LHM.  The results of this work would affect (at least) the research 
fields mentioned above on the very fundamental level.  

This paper is arranged as follows.  In Sec.2, we give a precise definitin of matter 
Hamiltonian and matter-EM field interaction, set up the simultaneous integral equations 
of microscopic nonlocal response, and, by applying LWA, extract the macroscopic components, 
which leads to a new macroscopic susceptibility $\chi_{\rm em}$.  The dispersion equation 
in terms of $\chi_{\rm em}$ is derived, and the conditions to reproduce the conventional 
one are studied.  Discussions are given in Sec.3 about (i) the interaction Hamiltonian 
in the form explicitly containing $\Vec{E}$ and $\Vec{B}$, and the condition for defining 
$\chi_{\rm e}$ and $\chi_{\rm m}$ separately, (ii) the comparison with other macroscopic theories, 
(iii) the comparison with the result of DBF eqs. for chiral symmetry, 
(iv) the consequence in the LHM studies, (v) the validity condition of LWA, and 
(vi) simplification of $\chi_{\rm em}$ with parameters.  In Sec.4, we summarize 
the important results. 

\section{Formulation} 

\subsection{Precise definition of matter, EM field and interaction} 

We start from the well-known Lagrangian for charged particles in EM field
\begin{equation}
\label{eqn:Lagrange}
  {\cal L} = \sum_{\ell}\ \{ \frac{1}{2} m_{\ell} v_{\ell}^2 
                         - e_{\ell} \phi(\Vec{r}_{\ell})
            + \frac{e_{\ell}}{c} \Vec{v}_{\ell}\cdot\Vec{A}(\Vec{r}_{\ell})\ \} 
         + \int {\rm d}\Vec{r}\  \frac{1}{8\pi}\{ 
    (\frac{1}{c} \frac{\partial \Vec{A}}{\partial t} 
    + \nabla \phi)^2 - (\nabla \times \Vec{A})^2 \}  \ ,
\end{equation}
where $\Vec{A}$ and $\phi$ are the vector and scalar potentials, and 
$e_{\ell}$, $\Vec{r}_{\ell}$ and $\Vec{v}_{\ell}$ are the charge, 
coordinate and velocity of the $\ell$-th particle, respectively.   
From the least action principle for this Lagrangian, we obtain the fundamental 
equations of the system, i.e. the microscopic M-eqs. and the Newton equation of 
each particle under Lorentz force.  

The microscopic M-eqs. are given in Coulomb gauge as 
($\nabla \cdot \Vec{A} = 0$), as   
\begin{equation}
\label{eqn:M-eqs-Coul}
 \nabla^2 \phi = - 4\pi\rho\ , \ \ \ \ \ 
 \frac{1}{c^2}\frac{\partial^2 \Vec{A}}{\partial t^2} - \nabla^2 \Vec{A} 
  = \frac{4\pi}{c} \Vec{J}_{\rm T} \ ,
\end{equation}
where $\Vec{J}_{\rm T}$ is the T component of 
$\Vec{J}$, i.e., $\nabla\cdot\Vec{J}_{\rm T} = 0$, and the charge density 
$\rho$ and current density $\Vec{J}$ are defined as 
\begin{eqnarray}
\label{eqn:rho}
  \rho(\Vec{r}) &=& \sum_{\ell} e_{\ell}\ \delta(\Vec{r} - \Vec{r}_{\ell})\ , \\ 
\label{eqn:c-densityJ}
  \Vec{J}(\Vec{r}) &=& \sum_{\ell} \ e_{\ell}\ \Vec{v}_{\ell}\ 
                              \delta(\Vec{r} - \Vec{r}_{\ell}) \ . 
\end{eqnarray}

The matter Hamiltonian 
including the interaction with EM field takes the form 
\begin{eqnarray}
\label{eqn:M-hamilton}
 H_{\rm M} &=& \sum_{\ell} \frac{1}{2m_{\ell}} 
            [\Vec{p}_{\ell} - \frac{e_{\ell}}{c}\Vec{A}(\Vec{r}_{\ell})]^2 
           + \frac{1}{2} \sum\sum_{\ell\neq\ell\ '\ \ \ \ } 
          \frac{e_{\ell}e_{\ell\ '}}{|\Vec{r}_{\ell} - \Vec{r}_{\ell\ '}|} \ .
\end{eqnarray}
in Coulomb gauge, where momentum $\Vec{p}$ is defined by eq.(\ref{eqn:B4x}). 
The $\Vec{A}$-linear part of this Hamiltonian 
\begin{equation}
\label{eqn:IA-int}
  H_{\rm intJ} = - \frac{1}{c}\ \int{\rm d}\Vec{r}\ 
                     \Vec{J}(\Vec{r})\cdot\Vec{A}(\Vec{r},t) .
\end{equation}
represents the linear light-matter interaction, and the $\Vec{A}$-independent 
part is the pure matter Hamiltonian. 

If one needs relativistic corrections, we add spin-orbit interaction, 
mass velocity term, and Darwin term to the matter Hamiltonian.  
Spin Zeeman term can be added to $H_{\rm intJ}$ because of their similar form as 
\begin{equation}
\label{eqn:spin-Z}
  H_{\rm sZ} = -\int{\rm d}\Vec{r}\ \Vec{M}_{\rm spin}(\Vec{r}) \cdot 
                \Vec{B}(\Vec{r}) 
     = - \frac{1}{c} \int{\rm d}\Vec{r}\ \Vec{J}_{\rm spin}(\Vec{r}) \cdot 
                \Vec{A}(\Vec{r})\ ,
\end{equation}
where the current density due to spin magnetization is 
\begin{equation}
\label{eqn:curr-d-spin}
  \Vec{J}_{\rm spin}(\Vec{r}) = c \nabla \times \Vec{M}_{\rm spin}(\Vec{r})
\end{equation}
and  
\begin{equation}
  \Vec{M}_{\rm spin}(\Vec{r}) = \sum_{\ell} \beta_{\ell} \Vec{s}_{\ell} \ 
                                \delta(\Vec{r} - \Vec{r}_{\ell})\ ,
\end{equation}
$\beta_{\ell} \Vec{s}_{\ell}$ being the spin magnetic moment of the $\ell$-th 
particle with spin $\Vec{s}_{\ell}$.   

Thus, the matter Hamiltonian is 
\begin{equation}
\label{eqn:B1}
 H_{0} = \sum_{\ell} \frac{\Vec{p}_{\ell}^2}{2m_{\ell}} 
           + \frac{1}{2} \sum\sum_{\ell\neq\ell\ '\ \ \ \ } 
          \frac{e_{\ell}e_{\ell\ '}}{|\Vec{r}_{\ell} - \Vec{r}_{\ell\ '}|} 
           + H_{\rm rc}\ ,
\end{equation}
including the relativistic correction $H_{\rm rc}$, 
and the matter-EM field interaction $H_{\rm int}$ is 
\begin{equation}
\label{eqn:B2}
  H_{\rm int} =- \frac{1}{c}\ \int{\rm d}\Vec{r}\ 
                     \Vec{I}(\Vec{r})\cdot\Vec{A}(\Vec{r},t) \ ,
\end{equation} 
where \begin{equation}
\label{eqn:cd-orb-spin}
  \Vec{I}(\Vec{r}) = \Vec{J}(\Vec{r}) + \Vec{J}_{\rm spin}(\Vec{r}) \ , 
\end{equation}
is the generalized current density operator, containing the contributions 
from both electric polarization and "spin and orbital" magnetization.  
The microscopic M-eqs. including this generalization is 
\begin{equation}
\label{eqn:M-eqs-I}
  \frac{1}{c^2}\frac{\partial^2 \Vec{A}}{\partial t^2} - \nabla^2 \Vec{A} 
  = \frac{4\pi}{c} \Vec{I}_{\rm T} \ .
\end{equation}

How to handle static EM field will depend on problems.  If we calculate 
static susceptibility, we will put the corresponding term in $H_{\rm int}$. 
If we are interested, e.g., in electron spin resonance or cyclotron resonance, 
however, we should put the static field part in $H_{0}$ and the time dependent 
part to $H_{\rm int}$. 

Another point of importance in considering EM response is how to treat L  
electric field induced in matter system.  It can arise from true charge density and 
electric polarization.  Let us assume for simplicity that matter system has no true 
charges ($\sum_{\ell} e_{\ell} = 0$).  
Its dynamically perturbed state is described by oscillating charge and current 
densities.   The polarization charge density $\rho_{\rm p}$ induces 
L electric field as 
\begin{equation}
\label{eqn:E-vs-rho}
  \Vec{E}_{\rm L}^{(\rm ind)}(\Vec{r}) = 
  - \nabla \int{\rm d}\Vec{r}' \frac{\rho_{\rm p}(\Vec{r}')}
              {|\Vec{r} - \Vec{r}'|}\ ,  
\end{equation}
which is a part of Maxwell field $\Vec{E}$. 
Its interaction energy with the polarization of the matter 
can be rewritten as the Coulomb interaction energy among the induced 
charge density, 
\begin{eqnarray}
\label{eqn:int-ind-cd}
 H_{\rm intL} &=& - \int{\rm d}\Vec{r}\ \Vec{P}(\Vec{r}) \cdot 
                 \Vec{E}_{\rm L}^{(\rm ind)}(\Vec{r})    \ , \\
\label{eqn:rho-rho-int}
             &=& \int{\rm d}\Vec{r} \int{\rm d}\Vec{r}'\ 
      \frac{\rho_{\rm p}(\Vec{r})\ \rho_{\rm p}(\Vec{r}')}{|\Vec{r}-\Vec{r}'|}\ .
\end{eqnarray}

This leads us to a  dichotomy, whether (A) we regard 
$\Vec{E}_{\rm L}^{(\rm ind)}$ as an internal field of 
matter, taking full Coulomb interaction into the matter Hamiltonian, or 
(B) we regard $\Vec{E}_{\rm L}^{(\rm ind)}$ as a part of the external source 
field to induce polarization, 
taking the Coulomb interaction (\ref{eqn:rho-rho-int}) as the interaction 
between the "external field" and matter polarization, i.e., omitting it 
from the Coulomb interaction energy of the matter.   
Since the omitted part of the Coulomb energy contributes to the energy 
difference between the T and L mode configurations, the scheme (A) 
contains the different T and L mode energies as the pole positions 
of susceptibility, while (B) reveals the effect of LT splitting, not in 
the susceptibility, but in response spectra.  Though the two schemes lead 
to the same observable result, there is a considerable difference in the 
intermediate steps.  

In this paper, we take the viewpoint (A).  Namely, we define the  
matter Hamiltonian as the sum of kinetic energy and (complete) 
Coulomb interaction of all the charged particles, and 
$\Vec{E}_{\rm L}^{(\rm ind)}$ is not considered as external field. 
Thus the Poisson equation contributes to the internal motion of matter, 
relevant to its quantum mechanical eigenstates, and only the second 
equation of (\ref{eqn:M-eqs-Coul}) represents the interaction between 
matter and EM field. (If we were to take the viewpoint (B), we would treat 
$H_{\rm int} + H_{\rm intL}$ as the interaction term, and omit 
(\ref{eqn:rho-rho-int}) from the Coulomb interaction of eq.(\ref
{eqn:M-hamilton}).   Using these matter Hamiltonian and matter-EM field 
interaction, we could calculate the time development of the operators 
$\Vec{I}, \Vec{P}, \Vec{M}$ according to the method in the next section.) 

Also we assume that the incident EM field is T mode, i.e., we do not 
consider the excitation by external charged particle.  Thus, in the 
Coulomb gauge, which is a natural choice for matter systems in 
non-relativistic regime, the matter Hamiltonian contains the full 
Coulomb interaction, and the external EM field is T-mode alone, i.e., 
$\Vec{A}$ or $\{\Vec{E}_{\rm T}\ {\rm and} \Vec{B}\}$.  
After determining the response, 
we can calculate $\Vec{E}_{\rm L}$ from the induced charge or 
current density.  (If we were to consider the excitation by external 
charged particle, which will be treated in a later publication, 
we would add an extra term 
\begin{equation}
\label{eqn:ext-ch-int}
 H_{\rm intL}^{\rm ext} = - \int{\rm d}\Vec{r}\ \Vec{P}(\Vec{r}) \cdot 
                 \Vec{E}_{\rm L}^{(\rm ext)}(\Vec{r})    \ , 
\end{equation}
to $H_{\rm int}$.  The external L field $\Vec{E}_{\rm L}^{(\rm ext)}$ and 
an external charge density $\rho(\Vec{r})_{\rm ext}$ are similarly related 
as in eq.(\ref{eqn:E-vs-rho}). In terms of the matter-EM field interaction, 
$H_{\rm int} + H_{\rm intL}^{\rm ext}$, we could calculate the induced current 
density, electric polarization, and magnetization as the expectation values 
of the corresponding operators. )

\subsection{Microscopic response}

In order to determine the microscopic response, we only need to have the microscopic   
constitutive equation between $\Vec{I}_{\rm T}$ and $\Vec{A}$, which is 
to be solved selfconsistently with eq.(\ref{eqn:M-eqs-I}).     
  
The microscopic calculation of induced current density can be found in various 
textbooks as e.g., \cite{AgranoSpd, Keldysh, Brenig, Cho-spr}, but the outline 
of the lowest order perturbation calculation is reproduced in Appendix B for the 
convenience of readers.  

The $\omega$ Fourier component of the induced current density 
is given in eq.(\ref{eqn:B8x}), or in an integral form as 
\begin{equation}
\label{eqn:micr-const}
  \Vec{I}(\Vec{r},\omega) = \int{\rm d}\Vec{r}' \chi(\Vec{r},\Vec{r}';\omega) 
         \cdot \Vec{A}(\Vec{r}',\omega)  
\end{equation}
where the integral kernel is the nonlocal susceptibility of the form 
\begin{equation}
\label{eqn:suscept}
  \chi(\Vec{r},\Vec{r}';\omega) = \frac{1}{c}\sum_{\nu} 
    \left[ 
      \bar{g}_{\nu}(\omega) \bar{\Vec{I}}_{0 \nu}(\Vec{r}) \bar{\Vec{I}}_{\nu 0} (\Vec{r}') 
        + \bar{h}_{\nu}(\omega) \bar{\Vec{I}}_{\nu 0} (\Vec{r}) \bar{\Vec{I}}_{0 \nu}(\Vec{r}')
    \,\right]
\end{equation}
and $\bar{g}_{\nu}(\omega)$ and $\bar{h}_{\nu}(\omega)$ are defined in eq.(\ref{eqn:bar-gh}). 
The operator $\bar{\Vec{I}}$ is the $\Vec{A}$ independent part of generalized current density 
(\ref{eqn:B5x}).  

The induced current density (\ref{eqn:B8x}) is a linear combination of the factors 
\{$F_{\mu\nu}$\}, defined in eq.(\ref{eqn:facF}), and the factors \{$F_{\mu\nu}$\} are 
determined by vector potential.  The selfconsistent solution of the coupled equations 
for current density and vector potential goes as follows.   

Using eq.(\ref{eqn:B8x}) as the source term of eq.(\ref{eqn:M-eqs-I}), 
we can solve the M-eq. and obtain $\Vec{A}(\Vec{r},\omega)$ as a linear 
combination of $\{F_{\nu 0}, F_{0\nu}\}$.   Substituting this solution in the 
r.h.s. of eq.(\ref{eqn:facF}), we obtain a set of simultaneous linear equations of 
$\{F_{\nu 0}, F_{0\nu}\}$, the solution of which determines selfconsistent 
$\Vec{A}$ and $\Vec{I}_{\rm T}$ (in the region inside the matter).  Once 
we have the selfconsistent $\Vec{A}$, we can use it to calculate $\Vec{I}_{\rm L}$ 
via eq.(\ref{eqn:B8x}), and $\Vec{A}$ at any point, e.g., response field outside 
a sample.  In this way the simultaneous equations of \{$F_{\mu\nu}$\} lead to   
all the information about the induced field and current density.   This is 
the main framework of microscopic nonlocal response theory mentioned in Sec.1.  

The nonlocality of the susceptibility is the characteristic feature of this theory. 
Within the extension of the relevant wave functions, an EM field applied at a 
microscopic point $\Vec{r}'$ can induce current density at different position $\Vec{r}$.  
Mathematically, the separable character of the integral kernel (susceptibility) 
plays an essential role to make this nonlocal scheme feasible in various 
manipulations, including the process of LWA to be discussed below.  

Another distinction from the macroscopic M-eqs. is the boundary conditions of EM field. 
To obtain the EM response in this microscopic scheme of nonlocal response, 
we do not need to consider the boundary conditions for EM field.  (Boundary conditions 
are already used to solve the eigenstates of matter, which determine the nonlocal 
susceptibility.)
Unique solution of the equations for \{$F_{\nu\mu}$\} is obtained only 
by requesting a given initial condition of EM field, i.e., incident field. 
This way of solution is in sharp contrast with the conventional macroscopic M-eqs., 
which need the boundary conditions for EM fields.  The various new features in 
microscopic response theory, suitable for the study of nanostructures in particular, 
are discussed in \cite{Cho-spr}. 
 
\subsection{Long wavelength approximation of microscopic response} 

  As mentioned in the introduction, the logically straightforward, reliable way of 
macroscopic averaging is to apply LWA to the microscopic response described in the 
previous subsection.   The validity condition of LWA depends on the system in 
consideration, so that it must be checked independently for each system. 
(See Sec.3.5 for more details.)   In this section, we discuss the 
genaral form of the macroscopic M-eqs., assuming LWA as a good approximation. 

If LWA is valid, the spatial variation of vector potential $\Vec{A}(\Vec{r},\omega)$ 
and induced current density $\Vec{I}(\Vec{r},\omega)$ will be weak in comparison with 
that of the matrix elements of the current density.   Thus, the variables $\Vec{A}$ 
and $\Vec{I}$ are represented by their long wavelength components alone.  The form of 
M-eqs. for $\Vec{A}$, eq.(\ref{eqn:M-eqs-I}), is kept unaltered under LWA.  In 
the Fourier representation, we have 
\begin{equation}
\label{eqn:M-eqsF}
 (-\frac{\omega^2}{c^2} + k^2) \tilde{\Vec{A}}(\Vec{k},\omega) 
      = \frac{4\pi}{c} \tilde{\Vec{I}}_{\rm T}(\Vec{k},\omega) 
\end{equation}
with the understanding that only small $k$ components have appreciable amplitudes.   
A similar expression holds for the constitutive equation (\ref{eqn:micr-const}) as 
\begin{equation}
 \label{eqn:c-density3}
 \tilde{\Vec{I}}(\Vec{k}, \omega) = \frac{1}{c}\sum_{\nu} 
    \left[  \bar{g}_{\nu}(\omega) \tilde{\Vec{I}}_{0 \nu}(\Vec{k}) F_{\nu 0}(\omega) 
        + \bar{h}_{\nu}(\omega) \tilde{\Vec{I}}_{\nu 0} (\Vec{k}) F_{0\nu}(\omega)
    \right]\ .
\end{equation}
Here also, only small $k$ components are considered to have appreciable amplitudes.   

The factor $F_{\mu\nu}(\omega)$ can be rewritten as 
\begin{equation}
 F_{\mu\nu}(\omega) = \sum_{\Vec{k}'} \tilde{\Vec{I}}_{\mu\nu}(-\Vec{k}')\cdot
                      \tilde{\Vec{A}}(\Vec{k}',\omega) \ , 
\end{equation}
which in general contains all the $\Vec{k}'$-components, but, if LWA is valid, 
only small $\Vec{k}'$s make the central contribution.   

For small $\Vec{k}$, we may take the first few terms of Taylor expansion 
(around $\Vec{r} = \bar{\Vec{r}}$) as  
\begin{equation}
 \tilde{\Vec{I}}_{\mu\nu}(\Vec{k}) = \frac{1}{V_{\rm n}}  
         \int{\rm d}\Vec{r} \exp[-i\Vec{k}\cdot\Vec{r}]\ \Vec{I}_{\mu\nu}(\Vec{r}) 
       = \frac{\exp(-i\Vec{k}\cdot\bar{\Vec{r}})}{V_{\rm n}}
         \ (\bar{\Vec{I}}_{\mu\nu} - i\Vec{k}\cdot\bar{\Vec{Q}}_{\mu\nu})
\end{equation}
where $\mu=0$ or $\nu=0$, $V_{\rm n}$ is the normalization volume, and
\begin{equation}
 \bar{\Vec{I}}_{\mu\nu} = \int{\rm d}\Vec{r}\ \Vec{I}_{\mu\nu}(\Vec{r})\ , \ \ \ 
 \bar{\Vec{Q}}_{\mu\nu} = \int{\rm d}\Vec{r}\ 
        (\Vec{r} - \bar{\Vec{r}})\ \Vec{I}_{\mu\nu}(\Vec{r}) 
\end{equation}
represent the moments of the electric dipole (E1) and magnetic dipole (M1) 
(plus electric quadrupole (E2)) transitions, respectively.  We choose $\bar{\Vec{r}}$ 
for each transition "$\mu \leftrightarrow \nu$" in such a way (e.g., at the center 
of impurity atom) that the moment $\bar{\Vec{Q}}$ represents its physical meaning 
correctly.  Then, we obtain the relation between 
$\tilde{\Vec{I}}(\Vec{k},\omega)$ and $\tilde{\Vec{A}}(\Vec{k}',\omega)$ as 
\begin{eqnarray}
  \tilde{\Vec{I}}(\Vec{k}, \omega) &=& \frac{1}{V_{n}c} \ 
        \sum_{\nu} \sum_{\Vec{k}'}\ {\rm e}^{i(\Vec{k}'-\Vec{k})\cdot\bar{\Vec{r}}} \bigl[
          \bar{g}_{\nu}(\omega) (\bar{\Vec{I}}_{0\nu} - i\Vec{k} \cdot \bar{\Vec{Q}}_{0\nu}) 
                          (\bar{\Vec{I}}_{\nu 0} + i\Vec{k}'\cdot \bar{\Vec{Q}}_{\nu 0})
     \nonumber \\ 
    &+&  \bar{h}_{\nu}(\omega) (\bar{\Vec{I}}_{\nu 0} - i\Vec{k} \cdot \bar{\Vec{Q}}_{\nu 0})       
                  (\bar{\Vec{I}}_{0\nu} + i\Vec{k}' \cdot \bar{\Vec{Q}}_{0\nu})  
            \bigr]\cdot  \tilde{\Vec{A}}(\Vec{k}',\omega) \ .
\end{eqnarray}

This expression of the induced current density in LWA 
still allows the mixing of different wave vector components, which corresponds 
to a macroscopic body without translational symmetry.   If the macroscopic 
medium obtained by LWA has a translational symmetry, as usually anticipated, 
we can keep only the $\Vec{k}' = \Vec{k}$ term in the above summation.   
In this case, we have  
\begin{equation}
\label{eqn:const-chi-em}
  \tilde{\Vec{I}}(\Vec{k}, \omega) = \chi_{\rm em}(\Vec{k}, \omega)\ \cdot  
                                   \ \tilde{\Vec{A}}(\Vec{k},\omega)
\end{equation}
where
\begin{eqnarray}
\label{eqn:chi-em}
  \chi_{\rm em}(\Vec{k},\omega) &=& \sum_{\nu}\ \frac{N_{\nu}}{c}\ \bigl[
          \bar{g}_{\nu}(\omega) (\bar{\Vec{I}}_{0\nu} - i\Vec{k} \cdot \bar{\Vec{Q}}_{0\nu}) 
          (\bar{\Vec{I}}_{\nu 0} + i\Vec{k} \cdot \bar{\Vec{Q}}_{\nu 0}) \nonumber  \\
        &+&  \bar{h}_{\nu}(\omega)  (\bar{\Vec{I}}_{\nu 0} - i\Vec{k} \cdot \bar{\Vec{Q}}_{\nu 0}) 
                     (\bar{\Vec{I}}_{0\nu} + i\Vec{k} \cdot \bar{\Vec{Q}}_{0\nu}) \bigr] \ .
\end{eqnarray}
Here, we have replaced the factor $1/V_{n}$ with the number density $N_{\nu}$ 
of localized states (of impurities, defects, etc.) corresponding to the transition 
$0 \leftrightarrow \nu$, and the summation over $\nu$ is to be taken only once 
for the same localised transitions at different sites. 

This is the general susceptibility of the present macroscopic scheme, and is the only 
susceptibility required to determine the complete (linear) response. 
Though this expression contains the both T and L components of $\tilde{\Vec{I}}$, only 
the T components are required for the selfconsistent solution with eq.(\ref{eqn:M-eqsF}). 
It contains 
the contributions from both electric (E1) and magnetic dipole (M1) transitions, 
together with their mixing terms.  It should be noted that the mixing terms remain 
nonvanishing in the case of chiral symmetry, where each excited state $|\nu>$ 
is active to both E1 and M1 transitions.  In this case, we cannot 
properly define $\chi_{\rm e}$ and $\chi_{\rm m}$, or $\epsilon$ and $\mu$.  
In the absence of chiral symmetry, on the other hand,  E1 and M1 characters 
are not mixed, so that the k-linear terms vanish, and the susceptibility turns out 
to be a sum of E1 and M1 types of terms.  This is the situation where we can use 
$\epsilon$ and $\mu$, and only in this case, the dispersion equation of the present 
formulation coincides with the conventional one in terms of $\epsilon$ and $\mu$, 
as will be discussed below in more detail.

\subsection{Dispersion equation of plane waves} 

The macroscopic constitutive equation obtained above by LWA is described by 
the susceptibility tensor $\chi_{\rm em}(\Vec{k},\omega)$. 
Substituting the expression of $\tilde{\Vec{I}}(\Vec{k}, \omega)$ in the source term 
$(4\pi/c) \tilde{\Vec{I}}$ of eq.(\ref{eqn:M-eqsF}), we get the equation 
\begin{equation}
  \left(\frac{c^2 k^2}{\omega^2} - 1\right) \tilde{A}_{\xi} = 
      \frac{4\pi c}{\omega^2} \sum_{\eta} (\chi_{\rm em})_{\xi\eta} \tilde{A}_{\eta}
\end{equation}
where $\xi, \eta$ are the two Cartesian coordinate axes perpendicular to $\Vec{k}$.  
The condition for the finite amplitude solution is the vanishing of the determinant 
of the coefficient ($2 \times 2$) matrix, i.e., 
\begin{equation}
\label{eqn:disp-cd}
  {\rm det}\left|\frac{c^2 k^2}{\omega^2} - 1 
   - \frac{4\pi c}{\omega^2}\ \chi_{\rm em}(\Vec{k}, \omega)\right| = 0\ .
\end{equation}
This is the dispersion relation in the present scheme of macroscopic M-eqs.  
It should be compared with the well known form of the dispersion relation 
in the traditional M-eqs. 
\begin{equation}
\label{eqn:disp-trad}
  {\rm det}\left|\frac{c^2 k^2}{\omega^2} - \epsilon \mu \right| = 0 \ . 
\end{equation}
Here also the T components of the tensors $\epsilon$ and $\mu$ should be inserted. 

Apparently, the two dispersion equations are different, because the contributions 
of electric and magnetic polarizations appear as a product in (\ref{eqn:disp-trad}), 
while in (\ref{eqn:disp-cd}) as a sum (including an interference term).   Moreover, 
the new result claims only one susceptibility, while there are two of them in the 
conventional formula.   In view of the possible mixing of E1 and M1 transitions in 
the case of chiral symmetry, $\epsilon$ and $\mu$ can have common poles, 
which leads to an unphyisical situation, i.e., the occurrence of second order poles 
in the product $\epsilon \mu$ in spite of the linear respose.  

In the absence of chiral symmetry, however, E1 and M1 (+ E2) transitions are 
grouped into different excited states.  Namely, there is no excited state $|\nu>$ 
making both of $\bar{\Vec{I}}_{\nu 0}$ and $\bar{\Vec{Q}}_{\nu 0}$ nonzero. 
In this case, we may divide the $\nu$ summation into two groups, so that we have 
\begin{equation}
  \frac{c}{\omega^2}\ \chi_{\rm em} = \bar{\chi}_{\rm e} + \bar{\chi}_{\rm m}\ ,
\end{equation}
where $\bar{\chi}_{\rm e}$ and $\bar{\chi}_{\rm m}$ are the partial summations 
over $\nu$ for E1 and M1 (+ E2) transitions, respectively, defined as 
\begin{eqnarray}
 \bar{\chi}_{\rm e} &=& \frac{1}{\omega^2} \sum_{\nu}\ N_{\nu}\ 
                  [\bar{g}_{\nu}(\omega) \bar{\Vec{I}}_{0\nu} \bar{\Vec{I}}_{\nu 0} 
                  +\bar{h}_{\nu}(\omega) \bar{\Vec{I}}_{\nu 0} \bar{\Vec{I}}_{0\nu}  ]\  
                        \ , \\
 \bar{\chi}_{\rm m} &=& \frac{k^2}{\omega^2} \sum_{\nu}\ N_{\nu}\ 
      [\bar{g}_{\nu}(\omega) (\hat{\Vec{k}} \cdot \bar{\Vec{Q}}_{0\nu})\ 
                      (\hat{\Vec{k}} \cdot \bar{\Vec{Q}}_{\nu 0}) 
       + \bar{h}_{\nu}(\omega)(\hat{\Vec{k}} \cdot \bar{\Vec{Q}}_{\nu 0})\ 
                      (\hat{\Vec{k}} \cdot \bar{\Vec{Q}}_{0\nu}) ]\
\end{eqnarray}
for a unit vector $\hat{\Vec{k}} = \Vec{k}/|\Vec{k}|$.  
In this case, the dispersion equation takes the form 
\begin{equation}
\label{eqn:disp-simple}
 {\rm det}\left|\frac{c^2k^2}{\omega^2} 
          - (1 + 4\pi\bar{\chi}_{\rm e} + 4\pi\bar{\chi}_{\rm m})\right| = 0 \ ,
\end{equation}
which should be compared with the traditional form of dispersion equation, 
(\ref{eqn:disp-trad})
\begin{equation}
\label{eqn:disp-tradx}
 {\rm det}\left|\frac{c^2k^2}{\omega^2} - (1 + 4\pi \chi_{\rm e})(1 + 4\pi \chi_{\rm m}) \right| = 0 \ .
\end{equation}
It appears that, even in this simplified case of the present formulation, 
the traditional form of dispersion equation cannot be recovered.  However, 
if we use, instead of $\chi_{\rm m}$, the more fundamental magnetic susceptibility 
$\chi_{\rm B}$ defined by $\Vec{M} = \chi_{\rm B} \Vec{B}$ (see 
the next section, Sec.3.1, for more details), we have $\mu = 1 + 
4\pi \chi_{\rm m} = (1 - 4\pi \chi_{\rm B})^{-1}$.  
Then, eq.(\ref{eqn:disp-tradx}) can be rewritten as 
\begin{equation}
\label{eqn:disp-trady}
 {\rm det}\left| \frac{c^2k^2}{\omega^2} - (1 + 4\pi \chi_{\rm e} + 4\pi \frac{c^2k^2}{\omega^2}  
           \chi_{\rm B})\right| = 0 \ ,
\end{equation}
where the contributions of E1 and M1 transitions appear as a sum and the M1 term 
contains the factor of $O(k^2)$, as in eq.(\ref{eqn:disp-simple}).   The relationship 
between ($\bar{\chi}_{\rm e}, \bar{\chi}_{\rm m}$) and ($\chi_{\rm e}, \chi_{\rm m}$) 
is obtained from the comparison of $O(k^0)$ and $O(k^2)$ terms of the 
two dispersion equations as 
\begin{equation} 
\label{eqn:chi-div}
  \chi_{\rm e} = \bar{\chi}_{\rm e} \ , \ \ \ 
  \chi_{\rm B} = (\omega/ck)^2 \bar{\chi}_{\rm m}  .
\end{equation}
In this way the equivalence of (\ref{eqn:disp-simple}) and (\ref{eqn:disp-tradx}) 
is demonstrated in the absence of chiral symmetry.   

The above argument gives the answer to all the problems 1) - 4) in the introduction: 
1) The electric and magnetic susceptibilities ($\chi_{\rm e}$ and $\chi_{\rm B}$) 
correspond to the first and second order terms, respectively, of the LWA expansion of 
the microscopic susceptibility $\chi(\Vec{r},\Vec{r}', \omega)$, in the absence of 
chirality.  Namely, they are two tensors derived from a single nonlocal 
susceptibility.  2), 3) Since we do not separate $\Vec{I}$ into different components, 
there arises no problem of non-uniqueness.  4) The (apparently) different wave number 
dependence of the two M1 transitions arises from the different means of description, 
i.e., $\chi_{\rm m}$ (or $\chi_{\rm B}$) in the case of spin resonance \cite{Slichter} 
and $\bar{\chi}_{\rm m}$ in the case of orbital M1 transition \cite{M1-tr}.

\section{Discussions} 

\subsection{Microscopic derivation of $\chi_{\rm e}$ and $\chi_{\rm m}$} 
To calculate induced electric and magnetic polarizations separately, we need the operator 
forms of these physical quantities.   It is known that the decomposition of $\Vec{J}$ 
into the current densities induced by electric and magnetic polarizations in the form of 
eq.(\ref{eqn:magpol}) is possible also in operator form.  If we use  
\begin{eqnarray}
\label{eqn:op-P}
 \Vec{P}(\Vec{r}) &=& \int_{0}^{1}{\rm d}u\ \sum_{\ell} e_{\ell} \Vec{r}_{\ell}\  
                   \delta(\Vec{r} - u \Vec{r}_{\ell})  \ , \\ 
\label{eqn:op-M}
 \Vec{M}(\Vec{r}) &=& \int_{0}^{1}u\ {\rm d}u\ \sum_{\ell} e_{\ell} \Vec{r}_{\ell}\  
                   \times \Vec{v}_{\ell}\ \delta(\Vec{r} - u \Vec{r}_{\ell}) \ , 
\end{eqnarray}
we can prove $\nabla\cdot\Vec{P} = -\rho$ and eq.(\ref{eqn:magpol}) for charge 
neutral systems (See Sec.IV.C of \cite{Cohen-Tann}). 

 Using these operator forms, we can calculate the induced electric and magnetic 
polarizations microscopically along the line of Appendix B.  
Then, the application of LWA to the microscopic constitutive equations 
gives the macroscopic (local) susceptibilities.  Such induced polaizations 
are given as functions of $\Vec{A}$, since the interaction $H_{\rm int}$ 
is given in terms of $\Vec{A}$, so that the susceptibilities do not correspond 
to $\chi_{\rm e}$ or $\chi_{\rm m}$.   This way of calculation 
leads to the equivalent result to that of Sec.2, because the calculated 
$\Vec{P}(\Vec{r})$ and $\Vec{M}(\Vec{r})$ gives the induced current density 
according to eq.(\ref{eqn:magpol}).  

 In order to introduce the variables of electric and magnetic field 
explicitly in the interaction Hamiltonian, the following transformation 
of Lagrangian is known to be useful.   Making use of the fact that the 
addition of a total time derivative of arbitrary function (of time and 
position) does not affect the least action principle of Lagrangian, 
we add the following term \cite{Cohen-Tann, PowZieWoo}
\begin{equation}
\label{eqn:add-F}
  F(t) = \frac{\rm d}{{\rm d}t}\ \frac{1}{c} \ \int{\rm d} \Vec{r} \ 
         \Vec{P} \cdot \Vec{A}
\end{equation} 
to the Lagrangian (\ref{eqn:Lagrange}).   
The combination of this term with $H_{\rm int}$ (\ref{eqn:IA-int}) leads to 
a new form of the interaction Hamiltonian as 
\begin{equation}
\label{eqn:PE-MB-int}
 H_{\rm int}' = -\int{\rm d}\Vec{r}\ \{\Vec{P}\cdot\Vec{E}_{\rm T}      
                      + \Vec{M}\cdot\Vec{B}\} \ ,
\end{equation}
where we have used partial integration and $\Vec{E}_{\rm T} = - (1/c) 
(\partial \Vec{A}/\partial t),\  \Vec{B} = \nabla \times \Vec{A}$ .     

The interaction Hamiltonian $H_{\rm int}'$ and the operator form of $\Vec{P}$ and 
$\Vec{M}$ appear to be appropriate for the calculation of electric 
and magnetic susceptibilities.  However, electric and magnetic fields cannot be applied  
independently for finite frequencies, and in chiral systems both of $\Vec{P}$ and 
$\Vec{M}$ can be induced by both $\Vec{E}$ and $\Vec{B}$.  This is not a situation to 
define electric and magnetic susceptibilities. 

Thus, the new form of interaction Hamiltonian does not guarantee the possibility of 
defining electric and magnetic susceptibilities in general.   In the absence of 
chirality, however, there is no mixing of E1 and M1 transitions, so that 
$\Vec{P}$ is induced by $\Vec{E}$ alone, and $\Vec{M}$ by $\Vec{B}$ alone, 
which allows the definition of electric and magnetic susceptibility. 
In this case, however, a perturbation calculation similar to the one used for 
$\tilde{\Vec{I}}(\Vec{k},\omega)$, (\ref{eqn:const-chi-em}), should give an induced 
magnetization in the form, not $\Vec{M} = \chi_{\rm m} \Vec{H}$, but 
$\Vec{M} = \chi_{\rm B} \Vec{B}$, since the interaction term (\ref{eqn:PE-MB-int}) 
is linear in $\Vec{B}$.   The poles of $\chi_{\rm B}$ correspond, similarly as in 
the case of $\chi_{\rm em}$, to the excitation energies of matter, i.e., magnetic 
excitations in this case.   Then, the usual definition $\Vec{B} = \Vec{H} + 4\pi \Vec{M}$ 
and $\Vec{M} = \chi_{\rm m} \Vec{H}$ lead to $\chi_{\rm m} = \chi_{\rm B} 
(1 - 4\pi \chi_{\rm B})^{-1}$ and $\mu = (1 - 4\pi \chi_{\rm B})^{-1}$.  
The last relation has 
provided an essential key to prove the equivalence of eq.(\ref{eqn:disp-simple}) and 
eq.(\ref{eqn:disp-tradx}).  The use of $\chi_{\rm B}$ is a logical consequence, but 
there are so many literatures using $\chi_{\rm m}$ on behalf of $\chi_{\rm B}$.  
The reason for it is difficult to find out from the literature of M-eqs.    
However, this paper is not the first one to claim the use of $\chi_{\rm B}$.  
See \cite{Brenig} for example.

\subsection{Comparison with other schemes of macroscopic M-eqs.}

There have been several proposals of single susceptibility schemes of macroscopic M-eqs. 
\cite{AgranoSpd, Agrano, Keldysh}, which are apparently motivated by the argument of Landau-Lifshitz
\cite{EMtext}, that, as frequency increases, magnetization tends to lose its physical meaning, 
i.e., it becomes meaningless to separate current density into the contributions of $\Vec{P}$ 
and $\Vec{M}$, as in eq.(\ref{eqn:magpol}).   This is very similar to our motivation, but 
our statement is one step stronger than theirs.  Namely, "for any frequency region, we should 
need only one vector field $\Vec{I}$ as in microscopic M-eqs., and it should be possible 
to write macroscopic M-eqs. without separating $\Vec{I}$ into the contributions of $\Vec{P}$ 
and $\Vec{M}$".   

In both of the proposals, they derive the microscopic forms of induced current density, 
from which one could derives conductivity and translate it into dielectric function.  
The calculation of conductivity is essentially equivalent to the content of Appendix B 
of this paper.  As to the translation of conductivity into dielectric function, it is 
a standard matter for the case of achiral symmetry, but it is not clear if one can do it 
also for chiral case. The authors of \cite{Agrano} call their scheme "$\Vec{E}, \Vec{D}, 
\Vec{B}$" approach.  In the same manner, we may call ours "(macroscopic) $\Vec{E}, 
\Vec{B}$" approach. In \cite{Keldysh}, they describe the case of dipole approximation, 
which corresponds to our $O(k^0)$ term of $\chi_{\rm em}$.  Our results in Sec.2 show the 
importance of the whole set of "$O(k^0), O(k^1), O(k^2)$" terms of the Taylor expansion  
to obtain the consistent picture of single suscptibility scheme of macroscopic M-eqs., 
which is free from the incompleteness problems 1) - 4) in Sec.1.  Both of \cite{AgranoSpd} 
and \cite{Keldysh} stress the distinction between the presence and absence of 
spatial dispersion within the macroscopic M-eqs., but we understand this simply as 
the different order terms of Taylor expansion within LWA.  In spite of the similarity in 
motivation and a part of formulas, neither of them show any systematic expansion 
of the microscopic constitutive equation to obtain a generalized form of macroscopic 
M-eqs.  In this sense, we could say that the present method and result reaches 
a more complete level of single susceptibility scheme. 
 
There are a few points which hinder the proper evaluation of these proposals. 
In \cite{Agrano}, they claim the existence of "one-to-one correspondence" between 
their scheme and the usual \{\Vec{E}, \Vec{B}, \Vec{D}, \Vec{H}\} framework.   
If this means the equivalence of the two schemes, their result is 
definitely different from ours.   Our result is, not equivalent to, but more general 
than \{\Vec{E}, \Vec{B}, \Vec{D}, \Vec{H}\} framework, based on the recognition 
of the latter's incompleteness problems and of the necessity to rationalize them.  
In \cite{Keldysh}, the microscopic matter Hamiltonian is described by the vector potential 
$\Vec{A}$ defined as $\nabla \times \Vec{A} = \Vec{H}$.  This should lead to the 
magnetic susceptibility $\chi_{\rm m}$ rather than to $\chi_{\rm B}$, yet "their 
magnetic susceptibility contributes to the single susceptibility a term of the type 
$(c^2k^2/\omega^2) \chi_{\rm m}$" (p.92, \cite{Keldysh}) as in eq.(\ref{eqn:disp-trady}).  
The formulas related with magnetic current density in Sec.1.5 of \cite{Keldysh}
\begin{equation}
\Vec{j}_{\rm m} = c\ \nabla\times \Vec{M} 
     = c\ \nabla\times \bar{\chi}_{\rm m}(\Vec{H} - 4\pi\Vec{M}) 
     \simeq  c\ \nabla\times \bar{\chi}_{\rm m}\Vec{H} \ . 
\end{equation}
is also difficult to understand.  It is not clear if their claim is equivalent to ours 
as to the necessity to prefer $\chi_{\rm B}$ over $\chi_{\rm m}$.  

There is a different approach to macroscopic M-eqs. by Nelson \cite{Nelson}, where he 
applies LWA to the Lagrangian of matter-EM field system, rewriting it into a "continuum"  
Lagrangian.   The explicit use of LWA in the mathematical treatment is a common feature 
to the present theory, but the physical meaning is quite different.   By the application 
of LWA to Lagrangian, the dynamics of matter is described only by the long wavelength (LW) 
components, i.e., the LW eigen modes of matter such as acoustic and optical phonons and 
excitons.   Thus the only contribution to susceptibility is made from the LW modes of 
matter, i.e. the susceptibility has poles only at the frequencies of these LW modes.  
Since all the dynamical variables of short wavelength components 
are eliminated by the LWA of Lagrangian, there is no chance for localized eigen states of 
the matter to contribute to susceptibility.  When we consider a problem, for example, of 
changing the refractive index of a material by adding impurities, the main change is caused 
by the localized excitations at the impurities.  But they cannot be taken into account in 
Nelson's treatment, because they are not LW modes of matter.   In contrast, our approach based 
on the LWA of microscopic constitutive equation takes all the contributions of the eigen modes 
of matter according to their weights in LWA, i.e., oscillator strengths.   Thus, our approach 
provides continuous relationship between microscopic and macroscopic descriptions of EM response, 
including the method to evaluate the validity condition of LWA (see Sec.3.5).

\subsection{Comparison with Drude-Born-Fedorov equation}
 
Chirality or gyrotropy of a system appears as the rotation of polarization plane 
\cite{chiral} and as the mixing effect between electric and magnetic dipole characters, 
e.g., of the excitons in CdS \cite{Hop-Th} and the coupled "Landau level - spin flip" 
transitions in GaAs \cite{Larocca}, and the Jones effect in atomic spectroscopy \cite{Budker}.  
It arises from the interference of electric and magnetic polarizations.  The description 
could be either microscopic \cite{Hop-Th}, \cite{Larocca}, \cite{Budker}, or macroscopic 
\cite{chiral}, depending on the problems.  The microscopic one is equivalent to the 
microscopic nonlocal response in Sec.2.2, and as macroscopic one we have two schemes, 
the present one in Sec.2.3 and 2.4, and that of DBF eqs. \cite{chiral}.  
  
In this subsection, we want to show that the two macroscopic schemes cannot be equivalent.  
For this purpose, we use the dispersion relations.  In our scheme, the dispersion relation 
is given as eq.(\ref{eqn:disp-cd}), where the terms in the determinant other than 
$c^2k^2/\omega^2$ is a sum of single poles except for $\omega = 0$.  This is a general 
aspect of linear response.  The dispersion equation for DBF eqs. is obtained by 
solving the M-eqs. $c \nabla \times \Vec{H} = \partial \Vec{D}/
\partial t, \ c \nabla \times \Vec{E} = - \partial \Vec{B}/\partial t$
and DBF constitutive eqs. (\ref{eqn:DBF}).  The condition for the existence of 
finite amplitude solution of the coupled linear equations for $\Vec{E}$ and $\Vec{H}$  
gives the dispersion relation.  In the simplified case where $\epsilon, \mu, \beta$ 
are scalars (with $\omega$ dependece), the dispersion equation is given as 
\begin{equation}
\label{eqn:chiraldisp}
 (\frac{ck}{\omega})^2 = \epsilon \mu \ 
            (1 \pm  \frac{\beta \omega}{c}\sqrt{\epsilon \mu})^{-2} \ .
\end{equation}
This is obviously different from the equation (\ref{eqn:disp-cd}) with respect   
to their pole structure on the r.h.s.   Since E1 and M1 characters of matter 
excitations are mixed in chiral materials, the successful trick in Sec.2.4 to divide 
$\chi_{\rm em}$ into $\chi_{\rm e}$ and $\chi_{\rm B}$ for achiral materials 
fails in this case. It is hopeless to rewrite $\chi_{\rm em}$ as a sum of E1, 
M1, and chiral components in accordance with DBF eqs., and to make the r.h.s. 
of eq.(\ref{eqn:chiraldisp}) a sum of single poles. 

This means that the DBF eqs. cannot be justified from a microscopic basis. 
Though the parameter $\beta$ can qualitatively describes the 
different phase velocities of right and left circularly polarized lights, 
we cannot obtain its quantum mechanical expression consistent with 
$\chi_{\rm em}$.  Thus, the use of the k-linear term of $\chi_{\rm em}$ is 
preferable to take care of the chirality, rather than the DBF eqs.  

\subsection{Influence on the studies of Left-Handed Materials (LHM)} 

According to the results of the previous section, Veselago's definition of LHM 
($\epsilon < 0, \mu < 0$) \cite{Veselago} needs to be revised because of the 
limited condition to allow the use of $\epsilon$ and $\mu$.  A more general 
definition would be "the occurrence of a dispersion branch with 
$v_{\rm ph} \times v_{\rm g} < 0$", where $v_{\rm ph}$ and $v_{\rm g}$ are 
phase and group velocities, respectively.  
The part of positive $v_{g}$ in the negative $k$ region is important because 
the plane wave on this part of the branch should be connected to the incident 
EM field via the boundary conditions at the interface.   When two (or more) 
plane waves exist in the medium at the frequency of an incident wave, the general 
criterion for the correct choice of the branch is that the connected wave in 
the medium should not diverge as the size of matter becomes larger.   From a 
general consideration of the dispersion curves in the complex ($\omega, k$) plane 
\cite{CuI-BC}, it is concluded that the choice of the branch with positive $v_{g}$ 
meets this criterion.  Physically, this means a simple fact that the plane wave with 
postive $v_{g}$ deays in the direction of its propagation.   Therefore, it applies to 
any dispersion curves of either right- or left-handed character. 

LHM can be made of either chiral or achiral materials.  Though it is recommended 
to use $\chi_{\rm em}$ in general, it is possible to use the conventional 
"$\epsilon, \mu$" scheme for achiral materials.  In this case, however, it is 
also required to assign the magnetic transition energies to the poles of, not 
$\chi_{\rm m}$, but $\chi_{\rm B}$.   Though the assignment to $\chi_{\rm B}$ 
is logically correct, that to $\chi_{\rm m}$ has been very frequently made 
in textbooks and papers.  In order to show the possibility of a serious difference 
due to this distinction, we give a measurable example below. 

A typical LHM behavior is expected in an achiral material when a magnetic 
transition occurs in the frequency region where 
$1 + (4\pi c/\omega^2)\bar{\chi}_{\rm e}\ (= \epsilon_{\rm b}) < 0$ .  
The two different assignments are 
\begin{eqnarray}
({\rm A})\ \ \  \chi_{\rm m} &=& \frac{b}{\omega_{0} - \omega - i0^+}, \ \\ 
({\rm B})\ \ \  \chi_{\rm B} &=& \frac{b}{\omega_{0} - \omega - i0^+},
\end{eqnarray}
where $b$ and $\omega_{0}$ are the strength and resonant frequency of the 
magnetic transition, respectively.  The corresponding dispersion relations are 
\begin{eqnarray}
({\rm A})\ \ \   \frac{c^2k^2}{\omega^2} &=& \epsilon_{\rm b}\ 
   (1 + \frac{4\pi b}{\omega_{0} - \omega - i0^+})\ ,  \\
({\rm B})\ \ \   \frac{c^2k^2}{\omega^2} &=& \epsilon_{\rm b}\ 
     (1 - \frac{4\pi b}{\omega_{0} - \omega - i0^+})^{-1}\ . 
\end{eqnarray}
The solution of this equation gives the dispersion relation $k = k(\omega)$ and refractive 
index $n = ck/\omega$.  For the positive $v_{g}$ branch of LHM behavior, $k$ and $n$ are negative. 
The reflection coefficient of a semi-infinite system for normal incidence is given as 
$R = |(n+1)/(n-1)|^2$ for a medium of LHM, with $n < 0$.   

\begin{figure}
 \begin{center}
  \includegraphics[width=7cm,clip]{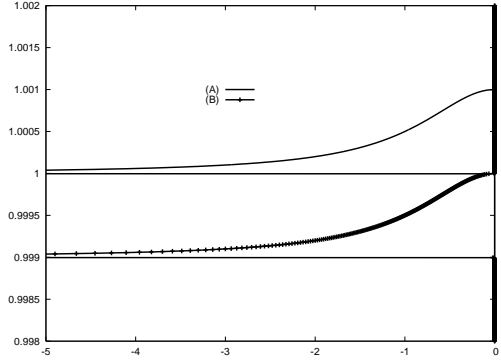}
 \end{center}
 \caption{Dispersion curves with $v_{\rm g} v_{\rm ph} < 0$.  The parameter values 
$\epsilon_{\rm b} = -1$ and $4\pi b / \omega_{0} = 0.001$ are used. The abscissa (k) 
and ordinate ($\omega$) are normalized by $\omega/c$ and $\omega_{0}$, respectively.
Two curves correspond to the cases (A) and (B) in the text.  The forms of the curves 
are quite similar, except for their positions with respect to the resonant frequency 
$\omega_{0}$. 
\label{fig:fig1} }
\end{figure}

Dispersion curve and reflectivity spectrum are shown in Fig.1 and Fig.2, respectively.  
We have a branch with $v_{\rm g} v_{\rm ph} < 0$ as expected, and this branch opens a window (a dip) 
in the total reflection spectrum ($R = 1$).   Since the value of $n$ along the dispersion curve changes 
continuously between 0 and $- \infty$, the dip of reflectivity can become as deep as 0 ($n=-1$).  
The remarkable point is the relative position of $\omega_{0}$ with respect to the reflectivity dip. 
In (A), $\omega_{0}$ occurs on the lower frequency end of the window, and in (B) 
on the higher frequency end.  This could be checked experimentally, via an appropriate model system, 
e.g., a magnetic resonance of well-defined impurities overlapping with an E1 type phonon resonance. 

\begin{figure}
 \begin{center}
  \includegraphics[width=7cm,clip]{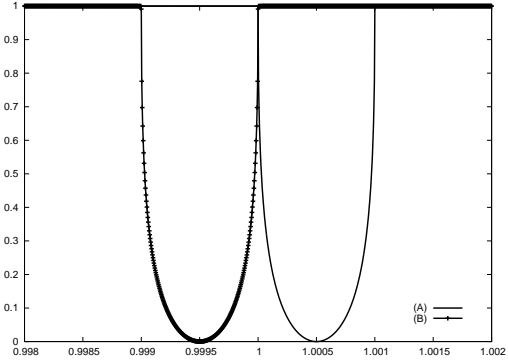}
 \end{center}
 \caption{Reflectance spectra of the two cases in Fig.1, showing transmission windows in the total 
reflection range.  Two curves are almost identical, except for their positions 
with respect to the resonant frequency $\omega/\omega_{0}=1$.}
\label{fig:fig2}
\end{figure}

\subsection{Validity of LWA}

In our derivation of the macroscopic M-eqs., we have just assumed the validity of LWA 
for the system in consideration, and applied LWA to the microscopic response.    
The validity condition of LWA is not provided by the macroscopic theory itself, but 
must be checked independently through the microscopic scheme described in Sec.2.2, 
where all the wavelength components are included in the response field.  
Only when the amplitudes of the short wavelength components are small enough 
in comperison with the LW component, LWA is a good approximation.   

The validity condition of LWA is closely connected with the resonant or non-resonant 
condition of the optical process in question. In resonant optical processes, 
only a few excited states will be resonant to incident frequency, and they will make 
the main contribution to the induced current density.  Thus, the current density 
acquires a characteristic microscopic spatial structure reflecting the quantum 
mechanical wave functions of the resonant excited states.   All the other non-resonant 
states make the contributions of more or less comparable amplitudes, so that their 
superposition will have no particular microscopic structure.   Therefore, non-resonant 
processes could generally be handled by macroscopic scheme, and resonant processes 
should mostly be treated by microscopic theory.   However, the resonant processes 
due to uniformly distributed well-localized impurity or defect states may well be 
treated by the macroscopic theory, as long as one considers the average properties 
of these localized states.   

In the case of resonant optical processes in confined systems, the breakdown of LWA can 
occur rather quickly as we increase the size of confinement starting from an atomic scale. 
(See sec.4.1 of \cite{Cho-spr}.) In this case, one should compare the confinement size, 
not with the wavelength of the resonant light in vacuum, but with that in matter which 
is shortened by the factor of background refractive index. 

\subsection{Simplification of $\chi_{\rm em}$ with parameters}

The merit of the conventional macroscopic M-eqs. lies in the description of matter 
in terms of a few material parameters.  Namely, for each material, one assumes certain 
simplified forms of $\epsilon$, $\mu$, and chiral admittance containing some parameters.  
Thereby one could consider resonant as well as nonresonant behavior of matter, so that 
one can handle a wide variety of EM responses.   These parameter values can be determined 
from the comparison between the solution of the M-eqs. and certain appropriate experiments.   

In this subsection, we discuss how far this kind of approach can be justifed from 
the results of this work.  While the conventional approach uses three susceptibilities 
$\epsilon$, $\mu$, and chiral admittance $\beta$ defined for different pairs of vector 
fields, our new result requires the use of a single susceptibility $\chi_{\rm em}$ 
between $\tilde{\Vec{I}}$ and $\tilde{\Vec{A}}$, which contains the three tensor terms 
of the order $O(k^0)$, $O(k^1)$ and $O(k^2)$ as 
\begin{equation}
  \chi_{\rm em} = \chi_{\rm e1} + ik\ \chi_{\rm chir}+ k^2\ \chi_{\rm m1} \ . 
\end{equation}
Each term on the r.h.s., to be obtained from eq.(\ref{eqn:chi-em}), consists of 
a summation over excited states $\nu$, which could be simplified as a constant (plus 
a few resonant terms).  This will reduce the number of freedom, facilitating the 
analysis of experimental results.   Whether or not the parameters in the simplified 
expressions can be treated as "free" should be judged from the original 
expressions.   In chiral materials, for example, E1 and M1 transitions mix each other, 
so that all the three terms have common single poles.  Therefore, if one keeps 
a pole in the simplification process, one should keep it for all of $\chi_{\rm e1}$, 
$\chi_{\rm chir}$ and $\chi_{\rm m1}$.

It would generally be safe to use $\chi_{\rm e1}$, $\chi_{\rm chir}$ and 
$\chi_{\rm m1}$ with common resonances. Especially, the models of LHM without 
inversion symmetry such as split ring resonators would generally need such a treatment. 
Only in nonchiral materials we can use "$\epsilon$ and $\mu$" with free parameters.  
But we should still be careful in ascribing a magnetic transition energy, 
not to the pole, but to the zero of $\mu$.

\section{Summary} 

The main points of our logical steps and important results are as follows. 
\begin{flushleft}
\underline{Logical steps}
\end{flushleft}
\begin{enumerate}
\item [1)] The derivation of the conventional macroscopic M-eqs. from microscopic basis 
is not complete because of the lack of uniqueness in dividing current density into the 
components arising from electric and magnetic polarizations, and of the apparent 
inconsistency with microscopic response about the number of constitutive equations. 
\item [2)] As a new method of macroscopic averaging, we apply LWA to the microscopic 
nonlocal constitutive equation (and the microscopic M-eqs.) without any other assumption. 
\item [3)] The whole derivation is made for the precisely defined Hamiltonian of charged 
particles interacting with oscillating EM field, where an explicit consideration is 
added about spins in the Hamiltonian and the current density operator. 
\end{enumerate}

\begin{flushleft}
\underline{Important results}
\end{flushleft}
\begin{enumerate}
\item[a)] The macroscopic M-eqs. obtained from the above mentioned procedure retain the same 
form as the microscopic M-eqs., written in terms of $\Vec{E}$ and $\Vec{B}$ (or $\Vec{A}$ 
and $\phi$), which require only one susceptibility tensor $\chi_{\rm em}$. 
\item[b)] This susceptibility tensor contains the contribution from both elecrtic and magnetic 
polarizations, together with their mutual interference, which is in generl not separable into 
different components of conventional type. 
\item [c)] Since it is not necessary to introduce $\Vec{P}$ and $\Vec{M}$, this scheme is 
free from the problems about uniqueness and consistency inherent to the conventional one.    
\item[d)] Matter with chiral symmetry, where E1 and M1 transitions are mixed, should be 
treated by the present framework.  The phenomenological scheme with additional chiral 
admittance of Drude-Born-Fedorov eq. cannot be reproduced from the microscopic theory.  
\item[e)] The dispersion equation is 
det$|(ck/\omega)^2 - 1 - (4\pi c/\omega^2) \chi_{\rm em}| = 0$ 
in general.  This can be reduced to the well known form 
det$|(ck/\omega)^2 - \epsilon \mu| = 0$ in 
the case of achiral symmetry.  Thereby, it is essential to use $\chi_{\rm B}$ 
(defined by $\Vec{M}=\chi_{\rm B}\Vec{B}$) and $\mu = (1 - 4\pi \chi_{\rm B})^{-1}$.  
\item [f)] The linear response coefficient of the magnetic transitions in nonchiral matter 
is not $\chi_{\rm m}$ (defined by $\Vec{M}=\chi_{\rm m}\Vec{H}$) but $\chi_{\rm B}$. 
An observable difference in these assignments is shown in connection with LHM behavior 
of a magnetic transition in the frequency region of $\epsilon < 0$. 
\item [g)] The susceptibilities $\chi_{\rm e}$ and $\chi_{\rm B}$ 
correspond to the first and second order terms, respectively, of the LWA expansion 
of the microscopic susceptibility $\chi(\Vec{r}, \Vec{r}', \omega)$ in achiral materials.  
\item [h)] Due to the restricted condition to use $\epsilon$ and $\mu$, a better definition of 
LHM would be the occurrence of dispersion branch with "(phase velocity)$\times$(group 
velocity) $< 0$". 
\item[i)] For the study of LHM, it is necessary to check whether or not a proposed model structure  
allows the use of $\epsilon$ and $\mu$, and to ascribe the resonant frequncy of magnetic 
excitation, not to $\chi_{\rm m}$, but to $\chi_{\rm B}$.  More generally, it is recommended to use 
$\chi_{\rm em}$. \vspace{10mm}
\end{enumerate}

\begin{flushleft}
\underline{Acknowledgments}\vspace{5mm}

  This work was supported in part by the Grant-in-Aid for Scientific Research (No.18510092) 
of the Ministry of Education, Culture, Sports, Sience and Technology of Japan.  The author 
is grateful to Prof. K. Shimoda, Prof. K. Ohtaka and Prof. M. Saitoh for useful discussions 
and comments. \vspace{10mm}

\underline{Appendix A. Fundamentals of M-eqs.} 
\end{flushleft}

Though the contents of this appendix is well-known, we give them explicitly 
in order to provide a basis to check all the formulas of this paper and 
to fix notations and units. 

Microscopic M-eqs. are the equations for two field variables, electric field 
$\Vec{E}$ and magnetic field (flux density) $\Vec{B}$ as 
\begin{equation}
\label{eqn:Mic-M-eqs}
  \nabla \cdot \Vec{E} = 4\pi \rho,\ \ \  
  \nabla \cdot \Vec{B} = 0, \ \ \ 
  \nabla \times \Vec{B} = \frac{4\pi}{c} \Vec{J} + 
   \frac{1}{c} \frac{\partial \Vec{E}}{\partial t},\ \ \  
  \nabla \times \Vec{E} = -\frac{1}{c} \frac{\partial \Vec{B}}{\partial t}\ .
\end{equation}
The source terms $\rho$ and $\Vec{J}$ are microscopic charge and current 
densities, respectively, satisfying the continuity equation, i.e., charge 
conservation law,  
\begin{equation}
\label{eqn:contin}
  \nabla \cdot \Vec{J} 
      + \frac{\partial \rho}{\partial t} = 0 \ . 
\end{equation}

These equations determine $\Vec{E}$ and $\Vec{B}$ for given $\rho$ and 
$\Vec{J}$.  To describe the EM response of matter, we need additional 
relationship to express $\rho$ and $\Vec{J}$ in terms of $\Vec{E}$ 
and $\Vec{B}$.  This kind of relationsip is called constitutive equation, 
and, in the present case, we generally need only one integral equation 
relating $\Vec{J}$ and "$\Vec{E}$ or $\Vec{B}$".  Since $\rho$ and 
$\Vec{J}$ are related via continuity equation, and $\Vec{E}$ and $\Vec{B}$ 
via Faraday's law, only one relationship between two vector fields is 
sufficient.  Thus, the number of required susceptibility tensor (integral 
kernel) is one.    

Macroscopic M-eqs. are used for the description of EM field in 
macroscopic matter, i.e., gases, liquids and solids, so that they employ 
macroscopic variables (usually electric and magnetic polarizations plus 
macroscopic current density) rather than the microscopic charge and current 
densities mentioned above as source terms of the equations.  
The central idea to derive the macroscopic from microscopic M-eqs. is the 
"macroscopic averaging of microscopic variables", which is made ''over 
a volume larger than atomic scale but smaller than light wavelength'' 
to extracts the macroscopic components of field and matter variables.  

The standard way to derive macroscopic M-eqs. from microscopic M-eqs. is 
to separate certain parts of charge and current densities as the contributions 
from electric and magneteic polarizations as 
\begin{eqnarray}
\label{eqn:pol}
 \rho &=& \rho_{\rm t} + \rho_{\rm p}\ , \ \ \ 
   \rho_{\rm p} = - \nabla \cdot \Vec{P}\ , \\
\label{eqn:magpol}
 \Vec{J} &=& \Vec{J}_{\rm c} + c\ \nabla \times \Vec{M} + 
             \frac{\partial \Vec{P}}{\partial t} \ ,
\end{eqnarray}
where $\rho_{\rm p}$ represents polarization charge density,  
$c\ \nabla \times \Vec{M}$ and $\partial \Vec{P}/\partial t$ 
the current densities due to magnetic and electric polarizations, respectively. 
The polarization charge density is defined for the part of neutral charge 
distribution, i.e, $\int{\rm d}\Vec{r}\ \rho_{\rm p} = 0$.  The remaining 
(non-neutral) part of the charge density contributes to $\rho_{\rm t}$. 
The true charge density $\rho_{\rm t}$ and the current density 
$\Vec{J}_{\rm c}$ caused by its motion satisfy the continuity equation 
of the form of eq.(\ref{eqn:contin}). 
In terms of the electric polarization $\Vec{P}$ and magnetic polarization 
$\Vec{M}$, we introduce the new field variables $\Vec{D}$ and $\Vec{H}$ as 
$\Vec{D} = \Vec{E} + 4\pi \Vec{P}$ and $\Vec{H} = \Vec{B} - 4\pi \Vec{M}$.  
Then, the microscopic M-eqs. are rewritten as  
\begin{equation}
\label{eqn:Mac-M-eqs}
 \nabla \cdot \Vec{D} = 4\pi \rho_{\rm t},\ \ \  
 \nabla \cdot \Vec{B} = 0,\ \ \ 
 \nabla \times \Vec{H} = \frac{4\pi}{c} \Vec{J}_{\rm c} + 
   \frac{1}{c} \frac{\partial \Vec{D}}{\partial t},\ \ \  
 \nabla \times \Vec{E} = -\frac{1}{c} \frac{\partial \Vec{B}}{\partial t}
\end{equation}
This set of equations are the macroscopic M-eqs. where all the field variables 
and source terms are macroscopically averaged quantities consisting of long 
wavelength components alone. 

The constitutive equations for macrsocopic M-eqs. are required for both 
$\Vec{P}$ and $\Vec{M}$ as $\Vec{P} = \chi_{\rm e} \Vec{E}, \ \ \Vec{M} = 
\chi_{\rm m} \Vec{H}$ in the case of linear response.  The material 
parameters $\chi_{\rm e}$ and $\chi_{\rm m}$ are called electric and 
magnetic susceptibility tensors, respectively, and they define the 
dielectric constant and magnetic permeability as    
$\epsilon = 1 + 4\pi \chi_{\rm e}, \ \ \mu = 1 + 4\pi \chi_{\rm m}\ \ 
(\Vec{D} = \epsilon \Vec{E}, \ \Vec{B} = \mu \Vec{H})$. 

There are low symmetry systems where electric and magnetic polarizations are 
mixed with one another, and the plane of polarization of a linearly polarized light 
rotates as it propagates.  This can occur either naturally or by the application 
of static electric and/or magnetic field.   This kind of system is called 
chiral, optically active and/or gyrotropic, and the phenomenological way of its 
description is known as Drude-Born-Fedorov (DBF) constitutive equations \cite{chiral}, 
\cite{Band}, which takes into account an additional susceptibility called chiral admittance 
$\beta$ as 
\begin{equation}
\label{eqn:DBF}
\Vec{D} = \epsilon (\Vec{E} + \beta \nabla\times \Vec{E}), 
\ \ \Vec{B} = \mu (\Vec{H} + \beta \nabla\times \Vec{H}) \ .
\end{equation}
In this case, we have two constitutive equations with three susceptibilities. \vspace{10mm}

\begin{flushleft}
\underline{Appendix B. Derivation of microscopic susceptibility} 
\end{flushleft}

Here we calculate the expectation value of current density operator with respect 
to the matter wave function at time $t$, $<\Psi(t)| \Vec{I}(\Vec{r}) |\Psi(t)>$.  
The time evolution of $\Psi$ is governed by the Schr\"odinger eq. 
$i\hbar \partial \Psi/\partial t = H_{0} + H_{\rm int}$, 
where the matter Hamiltonian $H_{0}$ and the matter-EM field interaction $H_{\rm int}$ 
are defined in eqs.(\ref{eqn:B1}) and (\ref{eqn:B2}), respectively.  
We neglect the $O(A^2)$ term in the interaction Hamiltonian, since we are interested 
in linear response.  

Using the interaction representation $\Psi = \exp(-iH_{0}\tau/\hbar) \tilde{\Psi}$, 
$i\hbar \partial \tilde{\Psi}/\partial t = H'(t) \tilde{\Psi}$, where 
\begin{equation}
\label{eqn:B4}
  H'(\tau) = \exp(iH_{0}\tau/\hbar)\ H_{\rm int} \ \exp(-iH_{0}\tau/\hbar) \ ,
\end{equation}
we can easily obtain the solution to the first order of $\Vec{A}$ as  
\begin{equation}
\label{eqn:B3}
 \tilde{\Psi}(t) = \tilde{\Psi}(-\infty) - 
   \frac{i}{\hbar}\ \int_{-\infty}^{t}{\rm d}\tau\ H'(\tau) \tilde{\Psi}(-\infty)                  
         + \cdots     \ . 
\end{equation}
The interaction $H_{\rm int}$ is adiabatically switched on in the remote past, where 
the matter state is assumed to be in the ground state $|0>$ of the Hamiltonian $H_{0}$, 
i.e., $\tilde{\Psi}(-\infty) = |0>$. 

The $\Vec{A}$ linear terms of the expectation value $<\Psi(t)| \Vec{I}(\Vec{r})|\Psi(t)>$ 
originate from those of $\Psi(t)$ and of the operator $\Vec{I}$.  The latter is in the velocity 
term in the orbital current density, eq.(\ref{eqn:c-densityJ}).  Since the velocity of 
each particle in the presence of vector potential is given as 
\begin{equation}
\label{eqn:B4x}
\Vec{v}_{\ell} = \frac{1}{m_{\ell}}[\Vec{p}_{\ell} 
              - \frac{e_{\ell}}{c} \Vec{A}(\Vec{r}_{\ell},t)] \ , 
\end{equation}
we can rewrite $\Vec{J}$ as 
\begin{equation}
\label{eqn:B5}
 \Vec{J}(\Vec{r}) = \sum_{\ell} \frac{e_{\ell}}{2 m_{\ell}}
              \left[\Vec{p}_{\ell} \delta(\Vec{r} - \Vec{r}_{\ell}) 
                  +    \delta(\Vec{r} - \Vec{r}_{\ell}) \Vec{p}_{\ell} \right]
             - \frac{1}{c} \hat{N}(\Vec{r}) \Vec{A}(\Vec{r},t) \ ,  
\end{equation}
where $\hat{N}(\Vec{r}) = \sum_{\ell} (e_{\ell}^2/m_{\ell})\  
\delta(\Vec{r} - \Vec{r}_{\ell})$ .

Let us separate the $\Vec{A}$ dependent part of $\Vec{I}$, i.e., 
the second term on the r.h.s. of eq.(\ref{eqn:B5}), as  
\begin{equation}
\label{eqn:B5x}
  \Vec{I} = \bar{\Vec{I}} - \frac{1}{c} \hat{N}(\Vec{r})\Vec{A}(\Vec{r},t) \ .
\end{equation}
Then, the $\Vec{A}$ linear part of the induced current density 
$<\Psi(t)| \Vec{I}(\Vec{r}) |\Psi(t)>$ is given as 
\begin{equation}
\label{eqn:B6}
 \Vec{I}(\Vec{r},t) = - \frac{1}{c} <0|\hat{N}(\Vec{r})|0> \Vec{A}(\Vec{r},t) 
               -  \frac{i}{\hbar}\ \int_{-\infty}^{t}{\rm d}\tau\ 
    <0|\left[\bar{\Vec{I}}(\Vec{r}) H'(\tau) - H'(\tau) \bar{\Vec{I}}(\Vec{r})\right]|0> 
\end{equation}
Using the eigenstates $\{|\nu>, |\mu> etc.\}$ of $H_{0}$, we can evaluate the matrix elements 
of $H'(\tau)$ as 
\begin{equation}
\label{eqn:B7}
 <\mu| H'(\tau) |\nu> = - \frac{1}{c} \exp[i(E_{\mu}-E_{\nu})\tau/\hbar] 
          \int{\rm d}\Vec{r}\ <\mu| \bar{\Vec{I}}(\Vec{r})|\nu> \cdot \Vec{A}(\Vec{r},t) \ .
\end{equation}
This form, together with the adiabatic switching factor $\exp(0^+\tau)$ in $H_{\rm int}$, 
where $0^+$ is an infinitesimally small positive number, allows us to carry out the time 
integral in eq.(\ref{eqn:B6}).  The $\omega$ Fourier component of $\Vec{I}(\Vec{r},t)$ 
gives the microscopic constitutive equation   
\begin{equation}
\label{eqn:B8}
 \Vec{I}(\Vec{r},\omega) = - \frac{1}{c} <0|\hat{N}(\Vec{r})|0>\ \Vec{A}(\Vec{r},\omega) 
      +  \frac{1}{c}\sum_{\nu} 
    \left[  g_{\nu}(\omega) \bar{\Vec{I}}_{0 \nu}(\Vec{r}) F_{\nu 0}(\omega) 
        + h_{\nu}(\omega) \bar{\Vec{I}}_{\nu 0} (\Vec{r}) F_{0\nu}(\omega)
    \right] \ ,
\end{equation}
where 
\begin{equation}
\label{eqn:facF}
  F_{\mu\nu}(\omega) = \int{\rm d}\Vec{r}\ 
             \bar{\Vec{I}}_{\mu\nu}(\Vec{r}) \cdot \Vec{A}(\Vec{r},\omega) \ \ \ 
             (\mu = 0, \ {\rm or}\ \nu = 0) \ ,
\end{equation}
and 
\begin{equation}
 g_{\nu}(\omega) = \frac{1}{E_{\nu 0} - \hbar \omega - i0^+} \ , \ \ \ 
 h_{\nu}(\omega) = \frac{1}{E_{\nu 0} + \hbar \omega + i0^+} \ .
\end{equation}

The first term on the r.h.s.of eq.(\ref{eqn:B8}) is the contribution of the 
ground state charge density.  This term is important in several situations, 
such as X-ray diffraction, low frequency conductivity, etc.  For resonant 
optical processes in nanostructures which is the most important playground 
for the microscopic resonse theory, however, the main contribution comes from 
a few resonant excited states and the first term makes one of very many 
nonresonant contributions which is altogether treated as a background 
susceptibility.   For the derivation of macroscopic M-eqs., which is the main 
purpose of this theory, it would be enough to retain the main part of the 
first term, $(-e^2 n_{0} /m_{0}c)\Vec{A}$, where $e$, $m_{0}$, and $n_{0}$ 
are the charge, free mass, and averaged ground state number density, 
respectively, of electrons in matter.   In this case, it is possible 
to rewrite the first term into a similar form as the second one, which 
results in a simple expression of the current density 
\begin{equation}
\label{eqn:B8x}
 \Vec{I}(\Vec{r},\omega) = \frac{1}{c}\sum_{\nu} 
    \left[  \bar{g}_{\nu}(\omega) \bar{\Vec{I}}_{0 \nu}(\Vec{r}) F_{\nu 0}(\omega) 
        + \bar{h}_{\nu}(\omega) \bar{\Vec{I}}_{\nu 0} (\Vec{r}) F_{0\nu}(\omega)
    \right] \ ,
\end{equation}
where 
\begin{equation}
\label{eqn:bar-gh}
  \bar{g}_{\nu}(\omega) = g_{\nu}(\omega) - \frac{1}{E_{\nu 0}}\ , \ \ \  
  \bar{h}_{\nu}(\omega) = h_{\nu}(\omega) - \frac{1}{E_{\nu 0}}\ , 
\end{equation}   
and $E_{\nu 0}$ is the matter excitation energy from the ground to the $\nu$-th 
excited state.   The terms with the factor $1/E_{\nu 0}$ are those arising from 
the first term eq.(\ref{eqn:B8}).  For this manipulation, we make use of the 
commutation relation between current density and dipole density operators and 
LWA is used for vector potential.  The details is given in Sec.2.4 and 2.5 of 
\cite{Cho-spr}.  

When matter state is described by ensemble more generally, e.g., a canonical 
ensemble at temperature $T$, one can use the perturbation 
expansion of the equation of motion of density matrix \cite{Keldysh}, \cite{Brenig}.  
(The result eq.(\ref{eqn:B8}) corresponds to the canonical ensemble of $T=0$ K.)    
Since the ($\Vec{r}, \Vec{r}'$) dependence of the microscopic susceptibility 
is also separable in this case, the process of applying LWA is very similar to 
that described in Sec.2.3.


\begin{thebibliography}{99}
\bibitem{Lorentz}
H. A. Lorentz, {\it The Theory of Electrons}, Teubner, Leipzig 1916
\bibitem{Cohen-Tann}
C. Cohen-Tannoudji, J. Dupont-Roc, and G. Grynberg, {\it Photons and Atoms}, 
Wiley Interscience, New York 1989, 
\bibitem{QED2}
{\it Selected papers on Quantum Electrodynamics}, ed. J. Schwinger, Dover New York 1958
\bibitem{EMtext} for example, 
L. D. Landau and E. M. Lifshitz, {\it Electromagnetics of Continuous Media}, Pergamon Press, 
Oxford 1960; 
J. H. van Vleck, {Theory of Electric and Magnetic Susceptibilities}, Oxford University 
Press, 1932 ; 
J. D. Jackson, {\it Classical Electrodynamics}, Third Ed., J. Wiley \& Sons, New York 1999
\bibitem{ChoICPS}
K. Cho, {\it Proc. 28th Int. Conf. on Physics of Semiconductors}, eds. W. Jantsch and F. Sch\"affler, 
APS 2007, p.1155
\bibitem{chiral}
P. Drude, {\it Lehrbuch der Optik}, Leipzig, 1900; M. Born, {\it Optik}, Heidelberg, 1972; 
F. I. Fedorov, Opt. Spectrosc. {\bf 6} (1959) 49; ibid. {\bf 6} (1959) 237
\bibitem{Band} e.g., 
Y. B. Band, {\it Light and Matter}, Wiley 2006, p. 142 
\bibitem{Slichter}
C. P. Slichter, {\it Principles of Magnetic Resonance}, Harper \& Row New York 1963 
\bibitem{M1-tr}
J.R.Oppenheimer, {Lectures on Electrodynamics}, Gordon and Breach, New York 1970, Chapter I, Sec.8
\bibitem{AgranoSpd}
V. M. Agranovich and V. L. Ginzburg, {\it Crystal optics with Spatial Dispersion, and Excitons}, 
Springer Verlag, Berlin Heidelberg 1984, Sec.6
\bibitem{Agrano}
V. M. Agranovich, Y. R. Shen, R. H. Baughman, and A. A. Zakhidov, Phys. Rev. B{\bf 69} (2004) 165112
\bibitem{Keldysh}
Yu. A. Il'inskii and L.V. Keldysh, {\it Electromagnetic Response of Material Media}, 
Plenum Press New York, 1994 
\bibitem{PC}
K. Inoue and K. Ohtaka, {\it Photonic Crytals} Springer Verlag, Heidelberg 2004
\bibitem{LHM}
S. A. Ramakrishna, Rep. Prog. Phys. {\bf 68} (2005) 449
\bibitem{NFO}
M. Ohtsu and H. Hori, {\it Near Field Nano-optics} Springer Verlag, Heidelberg 1999
\bibitem{Veselago}
V. G. Veselago, Soviet Phys. Uspekh {\bf 10} (1968) 509 
\bibitem{Brenig}
W. Brenig, {\it Statistical Theory of Heat}, Springer Verlag, Berlin, heidelberg, New York, 1989, 
pp.152
\bibitem{Cho-spr}
K. Cho, {\it Optical Response of Nanostructures: Microscopic Nonlocal Theory}, 
Springer Verlag, Heidelberg 2003; Errata, Web site of Springer Verlag for this book
\bibitem{PowZieWoo}
E. A. Power and S. Zienau, Philos. Trans. Roy. Soc. {\bf A251} (1959) 427; 
R. G. Woolley, Proc. Roy. Soc. Lond. {\bf A321} (1971) 557
\bibitem{Nelson}
D. F. Nelson: {\it Electric, Optic, and Acoustic Interactions in Dielectrics}, 
J. Wiley \& Sons Inc., New York 1979
\bibitem{Hop-Th}
J. J. Hopfield and D. G. Thomas, Phys. Rev. Lett, {\bf 4} (1960) 357
\bibitem{Larocca}
M. Dobrowolska, Y.-F. Chen, J. K. Furdyna, and S. Rodriguez, Phys. Rev. Lett. {\bf 51} (1983) 134
\bibitem{Budker}
D. Budker and J. E. Stalnaker, Phys. Rev. Lett., {\bf 91} (2003) 263901 
\bibitem{CuI-BC}
S. Suga, K. Cho, Y. Niji, J. C. Merle, and T. Sauder, Phys. Rev. B{\bf 22} (1980) 4931




\end{thebibliography}
\end{document}